\documentclass[preprint2]{aastex}

\newcommand{\lya}{Ly$\alpha$}
\newcommand{\ha}{H$\alpha$}

\newcommand\wha{$W_{\mathrm{H\alpha}}$}

\newcommand\fesclyc{$f_{\mathrm{esc,LyC}}$}
\newcommand\fion{$f_{\mathrm{esc,ion}}$}

\newcommand\llyc{$L_{\mathrm{LyC}}$}
\newcommand\lha{$L_{\mathrm{H}\alpha}$}

\newcommand{\esca}{erg/s/cm$^2$/\AA}

\newcommand{\oiii}{[O{\sc iii}]}

\shorttitle{Lyman Continuum escape from $z=2.2$ H-alpha emitters}
\shortauthors{Sandberg et al.}

\begin{document}


\title{Limits on Lyman Continuum escape from $z=2.2$ H-alpha emitting galaxies}









\author{A. Sandberg, G. {\"O}stlin, J. Melinder, A. Bik}
\affil{Department of Astronomy, Stockholm University, Oskar Klein Center, SE-106 91 Stockholm, Sweden}
\email{sandberg@astro.su.se}

\and

\author{L. Guaita}
\affil{NAF-Osservatorio Astronomico di Roma, Via Frascati 33, I-00040, Monte Porzio Catone, Italy}


\begin{abstract}
The leakage of Lyman continuum photons from star forming galaxies is an elusive parameter. When observed, it provides a wealth of information on star formation in galaxies and the geometry of the interstellar medium, and puts constraints on the role of star forming galaxies in the reionization of the universe.
\ha -selected galaxies at $z\sim2$ trace the highest star formation population at the peak of cosmic star formation history, providing a base for directly measuring Lyman continuum escape. Here we present this method, and highlight its benefits as well as caveats. We also use the method on 10 \ha\ emitters in the Chandra Deep Field South at $z=2.2$, also imaged with the Hubble Space Telescope in the ultraviolet. We find no individual Lyman continuum detections, and our stack puts a 5$\sigma$ upper limit on the average absolute escape fraction of $<$24\%, consistent with similar studies. 
With future planned observations, the sample sizes would rapidly increase and the method presented here would provide very robust constraints on the escape fraction. 
\end{abstract}



\section{Introduction}

As galaxies undergo episodes of star formation, a fraction of the newly formed stars will be massive O- and B-type stars, producing hard ultraviolet (UV) radiation with $\lambda < 912$ \AA, capable of ionizing hydrogen. Unless absorbed by dust, these Lyman Continuum (Ly C) photons are absorbed by neutral hydrogen and re-emitted as hydrogen recombination lines (e.g. \lya , \ha ), while the fraction of ionizing radiation that escapes cleanly (\fion ) is a difficult parameter to observe directly. This fraction relates to a multitude of different topics related to galaxy formation; it affects the normalization of star formation rates inferred from recombination lines, and could provide detailed information on the geometry of the interstellar medium (ISM) in galaxies, as well as the diffuse UV background permeating the universe. Perhaps most importantly, ionizing radiation escaping from the first galaxies is thought to have been the main force driving the reionization of the universe at $z>6$ \citep[e.g.][]{robertson-2013,bouwens-2015}. 
These recent models of the reionization of the universe contain a large uncertainty in that \fion\ for star-forming galaxies at high redshift is not well known. Detailed measurements of this quantity, even if they are only upper limits, can put strong constraints on the parameters of these models and thus help us understand if the first stars and galaxies could provide enough ionizing photons to drive reionization on their own. 

Several attempts have been made to measure the relative fluxes on the blue and red side of the Lyman limit (912 \AA) in $z>3$ Lyman Break (LBG) and \lya\ emitting (LAE) selected galaxies using ground-based photometry and spectroscopy \citep{steidel-2001,iwata-2009,vanzella-2012,nestor-2013,grazian-2015}. However, these methods have several caveats; 
1) The production rate of ionizing photons ($Q$) is a strong function of the star formation history (SFH), which is difficult to constrain with rest frame near-UV observations or SED analysis.  
2) Even if the SFH is known, there are theoretical uncertainties from stellar evolution models (IMF differences, contributions from high-mass binaries, etc.) that affect the predicted value of $Q$. For example, $Q$ can vary as much as a factor of 5 depending on the rotation rate of massive stars \citep{leitherer-2014,maeder-meynet2012,vazquez-2007}. 
3) As is known from studies of the cosmic star formation rate (SFR), the presence of dust means that UV observations only probe a small fraction of the total star formation. Hence, a UV selected sample is not tracing the cosmic $Q$ in a fair way.

Another selection method for measuring \fion\ is to use \ha\ selected galaxies, since \ha\ is directly related to the star formation and is unaffected by caveats 1 and 2: The \ha\ emissivity is directly proportional to $Q$, where the only uncertainty comes from dust extinction. However, the fraction lost to dust is much smaller than in the UV: \ha\ selected galaxies typically suffer about one magnitude of \ha\ extinction 
\citep{sobral-2013,hayes-2010,reddy-2008}. An \ha\ selected sample will thus contain proportinally more dusty galaxies (which are less likely to show Ly C leakage), but they will be a less biased measure of the total cosmic star formation, and hence the global escape fraction. We note that for very high escape fractions, the \ha\ luminosity would be lower, since less nebular recombinations would take place. However, the escape fraction is not expected to be that high and this effect would probably not affect this type of study.

Correcting the UV flux requires detailed knowledge of the UV extinction law, where differences between extinction laws in different environments can lead to very large correction factors with uncertainties of over an order of magnitude. An \ha\ selected sample also needs extinction corrections, but the uncertainties are considerably smaller. In principle, other star formation rate tracers could be used, such as mid-IR emission. However, these are based on empirical relations, trace star formation on different time scales and are subject to similar uncertainties in stellar evolution theory as the UV method, and are not sensitive enough for the faintest galaxies. 

Ideally, one would measure \fion\ close to the end of the epoch of reionization, but due to the opacity of the Earth's atmosphere at longer wavelengths, \ha\ can be efficiently surveyed from the ground only up to $z\sim2.4$. At these redshifts the Lyman limit is in the UV and requires space-based observations.  At higher $z$ ($>3$) the LyC can be probed from the ground,
but at $z>2$ the opacity of the IGM due to intervening absorption systems increases rapidly with $z$. Correcting for this in small to moderate samples is intrinsically difficult as the amount of intervening material varies from sightline to sightline. At $z=2$ the average transmitted LyC fraction at $\sim$ 900 \AA is $\sim 75\%$ according to the latest estimates \citep{inoue-2014}, requiring a correction factor of 1.3. At $z=4$ the same correction factor is 5, and very sensitive to cosmic variance. 

Moreover, ground based rest frame UV surveys typically have seeing $\sim 1\arcsec$ and are prone to spurious results from low redshift interlopers. Indeed, most ground based detections have on closer inspection been found to be interlopers \citep[e.g.][who argue that HST imaging and deep spectroscopy with 8 meter class telescopes are required to confirm or reject LyC candidates]{vanzella-2012,siana-2015}.

An \ha\ selected sample of galaxies in fields with deep HST UV and multi-wavelength imaging is thus advantageous for probing \fion\ at $z\sim2$, when the universe was only $\sim$20\% of its current age and peaking in star formation \citep[e.g.][and references therein]{madau-dickinson2014}. It was these considerations that lead us to start an observing campaign, targeting fields with deep UV observations with \ha\ narrowband imaging. The current paper represents a small pilot study, demonstrating the strengths of the project and the results that it may bring in the future. 

\section{Matching \ha\ emitters to WFC3/F225W data}

The \ha\ observations were presented in \citet{hayes-2010}. This was a matched study of \ha\ and \lya\ emitters at $z = 2.2$ in the GOODS-South field. \ha\ emitters were selected by a narrowband excess, requiring \wha\ $>$ 20 \AA\ as measured by the difference in color between the continuum $K_s$ and the narrow $NB2090$ filter. Sources were also required to have a narrowband flux a factor $\Sigma = 5$ greater than the noise in the continuum image, and a photometric redshift consistent with $z=2.2$. The final sample consisted of 55 \ha\ emitter candidates. 

The Hubble Ultra Deep Field (UDF), which is part of GOODS-South, was observed with HST/WFC3 in the UV, using the F225W, F275W and F336W filters, as part of the UVUDF HST treasury program \citep{teplitz-2013,rafelski-2015}. At $z=2.2$, any pure LyC emission will be only in the F225W filter, whereas F275W would contain also non-ionizing continuum emission. 

\begin{figure}
\plotone{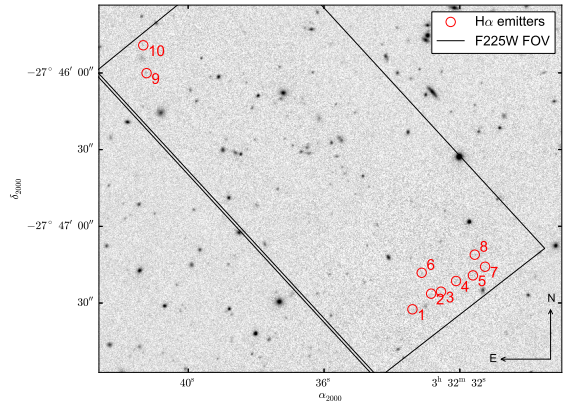}
\caption{Overview of \ha\ emitters from VLT/Hawk-I NB2090 imaging presented in \citet{hayes-2010} overlapping with the UV observations. \ha\ emitters are marked with red circles.  
The black outline shows the footprint of the HST WFC3/F225W UV observations of \citet{teplitz-2013}. We show only the portion which contains overlapping \ha\ emitters; the other chip (to the southeast) contains no sources.}\label{fig:FOV}
\end{figure}

Figure~\ref{fig:FOV} shows a cutout of the NB2090 \ha\ image, with our \ha\ candidates marked, and the overlap with the F225W observations. We originally find 14 \ha\ sources within the UV footprint. One source is located at the very edge of the WFC3 detector and is therefore not included. We find one of our remaining sources has a spectroscopic redshift in the 3dHST catalog  \citep{brammer-2012,skelton-2014} identifying it as an AGN at $z = 3.193$, which is perfectly consistent with \oiii\ 4959 emission falling into the NB2090 filter. 
We then look for individual sources by eye in the UV image that match the positions of our \ha\ emitters. We find matches for two of our 12 sources within an arcsecond of the \ha\ central coordinates. These sources have clear counterparts in the multiwavelength 3dHST catalogs, and their photometric redshifts are $z \sim 1.43$ and $z \sim 1.06$, with small uncertainties. 
Examining the SEDs of these two objects by eye, they both show very clear breaks consistent with the estimated photometric redshifts. These are not spectroscopically confirmed redshifts, but with their high F225W-F275W colors we see these sources as very probable foreground contaminating sources unrelated to any possible $z=2.2$ \ha\ or LyC emission. We have excluded these two sources from our stacking analysis so as not to contaminate the sample. Our final sample thus consists of 10 \ha\ emitter candidates at $z=2.2$ (see Figure~\ref{fig:FOV}).
 
\section{Stacked UV data and limits on \fion}

\begin{figure}
\plotone{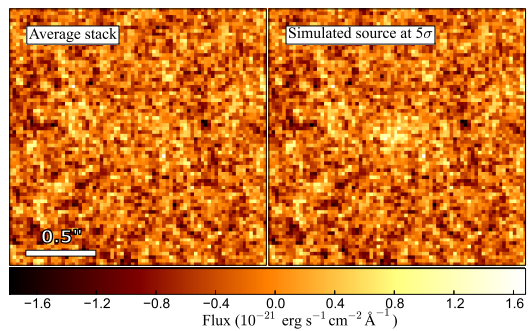}
\caption{Left: Average stack of WFC3/F225W cutouts, centred on the 10 \ha\ emitter candidates' coordinates. Right: Simulated ten stacked sources, each with integrated brightness of $1.49 \cdot 10^{-19}$ \esca , corresponding to the 5$\sigma$ detection limit, added to the stacked data in the left panel.}\label{fig:UVstack}
\end{figure}

We make cutouts of the WFC3/F225W image, centered on the coordinates of the 10 remaining photometric narrowband \ha\ detections. We then stack these cutouts, both using the median and the average. Note, however, that any non-isotropic signal will be effectively removed in a median stack, and we show only the average stack here (see Fig.~\ref{fig:UVstack}). 
Our final stacks contain no significant signal at the 1$\sigma$ level. 

In order to estimate an upper limit on \fion , we must make a few assumptions. Following \citet{kennicutt-1998}, we can express the H-alpha luminosity in terms of the number of ionizing photons produced per second ($Q$), taking into account the escaping photons:
$L_{\mathrm{H}\alpha} = 1.4 \cdot 10^{-12} \times Q \times (1 - f_{\mathrm{esc,LyC}})~\mathrm{erg/s}$.
In order to estimate the relation between $Q$, \fesclyc , and the LyC luminosity (\llyc ), we use a Starburst 99 spectral model \citep{leitherer-1999}, with a Salpeter IMF, metallicity $Z = 0.004$ and 10 Myr of constant star formation. 

The F225W filter is slightly asymmetric, and samples LyC at $z=2.2$ mostly in a range from 660 to 815 \AA , with a long red tail extending up to $\sim 900$ \AA. We therefore do not observe the total LyC luminosity, but rather the portion that falls into the filter. We denote this luminosity density by $L_{\lambda 740}$, and by convolving the LyC flux of our model with the F225W transmission curve, we arrive at an expression for the observed LyC luminosity density; 
$L_{\lambda 740} = 5.5 \cdot 10^{-14} \times Q \times f_{\mathrm{esc,LyC}}~\mathrm{erg/s/}$\AA. 
Note, however, that we also experiment with other spectral synthesis parameters to get a handle on how the result changes based on these assumptions. Any parameter that could alter the shape shortward of the Lyman break will have an effect on the scaling between $Q$ and $L_{\lambda 740}$, and the factor is also dependent on how representative this window is of the total \llyc . For an age of 1 Myr (3 Myr), we find the scaling factor to be $4.8 \cdot 10^{-14}$ ($5.2 \cdot 10^{-14}$). Increasing the metallicity to $Z = 0.02$ raises the scaling factor by about 12\%, and truncating the IMF at $30~\mathrm{M}_{\odot}$ (instead of 100 $\mathrm{M}_{\odot}$) increases it by 20\%. Changing the star formation history to a single burst has a large effect; the scaling factors are $4.8 \cdot 10^{-14}$, $6.2 \cdot 10^{-14}$ and $7.9 \cdot 10^{-14}$ for an age of 1, 3 and 10 Myr respectively. We must also factor in the possible effects of rotation on the production of ionizing photons, as described in \citet{leitherer-2014}. Judging from their models, the total $Q$ might increase with as much as 40\% if all stars rotate with 40\% of the break-up velocity. We are trying to constrain the upper limit on the LyC escape and so the smallest values of the scaling factor are the most important, since they allow for the largest $f_{\mathrm{esc,LyC}}$. 

Taking these considerations into account, we choose to use a more conservative scaling of
$L_{\lambda 740} = 4 \cdot 10^{-14} \times Q \times f_{\mathrm{esc,LyC}}~\mathrm{erg/s/}$\AA , 
and we arrive at a relation between \lha, \llyc\ and the escape fraction;
$f_{\mathrm{esc,LyC}} = \big[ 1 + \big( 35 \frac{L_{\lambda 740}}{L_{\mathrm{H}\alpha}} \big)^{-1} \big]^{-1}$

The limiting magnitude at 5$\sigma$ in the F225W filter is $\sim$ 27.8 \citep{teplitz-2013}, which corresponds to $1.49 \cdot 10^{-19}$ \esca . This is consistent with the noise that we measure in the images as well (see right panel of Figure~\ref{fig:UVstack}). The Double-Blind \ha\ sources are faint, with a mean \ha\ luminosity of \lha = $4.6 \times 10^{41}$ erg/s.  

For estimating the transmissivity through the IGM, we use a recent model \citep{inoue-2014} at z=2.2 and convolve it with the F225W response curve. In this way, we estimate the transmissivity to 42\%, and we estimate the escape fraction detection limit to be $\sim$50\% for a single source. If we assume all of our 10 sources to be real \ha\ emitters at $z = 2.2$, the final upper limit on \fion\ is $\sim$24\%.

\section{Conclusions and outlook}

This study is currently strongly restricted due to the small fields probed, especially in the UV. For larger fields, the number of bright \ha\ emitters available would increase greatly. These brighter galaxies produce intrinsically more ionizing photons (higher values of $Q$), which immediately puts stronger constraints on the detectability of LyC escape. The improved statistics of larger surveys also mitigate viewing angle and cosmic variance effects which can otherwise seriously bias the results. As is clearly visible in Figure~\ref{fig:FOV}, the sample of \ha\ emitters we present here is very strongly affected by cosmic variance and may not be representative of the average $z\sim2$ galaxy population. 

GOODS-South is currently being imaged further with HST in the F275W filter (P.I. Oesch), 
which will greatly increase the footprint of the existing \citet{teplitz-2013,rafelski-2015} F275W data. The \ha\ data that we consider here is at too low redshift for using F275W, allowing non-ionizing continuum photons to contaminate the photometry. However, a similar near-infrared study with narrowbands at slightly longer wavelengths (corresponding to $z\gtrsim2.38$) would allow a similar study to that presented here, including much more UV data. 

Other notable imaging campaigns with the HST in the UV involve the six Frontier Fields and Abell\,1689, which have already been imaged in F275W, and more data are on their way (P.I. Siana). We hope to observe these fields with extensive near-infrared narrowband imaging in the near future.

Our results can be compared with the recent findings of \citet{mostardi-2015}, who find a single LyC emitting galaxy with absolute escape fraction $14\pm7\%$ in a sample of 16 candidates at $z\sim3$. \citet{grazian-2015} study 45 galaxies at $z\sim3.3$, revealing only two Ly C leaking candidates, for a total average relative escape fraction below 2\%. These studies suffer from the caveats mentioned in the introduction, but do indicate that bright galaxies seem unlikely to have contributed enough to reionization on their own, without invoking rapid changes in their physical properties until $z\gtrsim6$. 

Our stacked 24\% limit on the absolute escape fraction assumes that all 10 candidates are real \ha\ emitting galaxies at $z=2.2$, but they are intrinsically faint and not spectroscopically confirmed. Note, however, that our average \ha\ luminosity assumes no dust extinction. A typical dust extinction of 1 magnitude \citep{sobral-2013,hayes-2010,reddy-2008} would raise the intrinsic median \lha\ to $\sim 10^{42}$ erg/s, and lower the stacked limit to $\sim$ 13\%. At the same time, LyC escape is very likely to be non-isotropic (due to outflows and/or low covering fractions along certain sightlines), which can still allow for high escape fractions in individual objects. However, the cosmologically interesting quantity is of course the average escape fraction. 

As pointed out by e.g. \citet{bouwens-2015}, a galactic relative escape fraction in the range 5 - 40\% might be enough for galaxies to reionize the universe, although the exact details are highly model dependent. There are of course uncertaintes in the possible evolution of the escape fraction from $z=6-8$ until $z=2$, but as we have mentioned, the escape fraction can only be reliably estimated at these lower redshifts. 



\acknowledgments

This work was supported by the Swedish Research Council and the Swedish National Space Board.



{\it Facilities:} \facility{HST (WFC3)}, \facility {ESO/VLT (Hawk-I)}.

\end{document}